# On chip, multifunctional quantum sensing using single spins in a van der Waals crystal

James Liddle-Wesolowski[1,2], Konosuke Shimazaki[1,2], Jiyun Kim[1, 2], Benjamin Whitefield[1,2], Kenji Watanabe[3], Takashi Taniguchi[4], Mehran Kianinia[1,2], and Igor Aharonovich[1,2]

[1] School of Mathematical and Physical Sciences, University of Technology Sydney, Ultimo, New South Wales 2007, Australia

[2] ARC Centre of Excellence for Transformative Meta-Optical Systems, University of Technology Sydney, Ultimo, New South Wales 2007, Australia

[3] Research Center for Electronic and Optical Materials, National Institute for Materials Science, 1-1 Namiki, Tsukuba, 305-0044, Japan

[4] International Center for Materials Nanoarchitectonics, National Institute for Materials Science, 1-1 Namiki, Tsukuba, 305-0044, Japan

**Abstract**

*Nanoscale thermometry and magnetometry are in high demand across a wide range of scientific and technological applications. In this context, optically addressable spins in solids have emerged at the forefront of on-chip quantum sensing. However, simultaneous quantum sensing of multiple parameters (e.g., temperature and magnetic field) using the same spin sensor remains challenging due to cross-sensitivity to multiple physical quantities. Here, we demonstrate independent dual sensing of temperature and magnetic field using single quantum emitters in hexagonal boron nitride (hBN). We experimentally verify the independent response of the zero-phonon line (ZPL) position to temperature and of optically detected magnetic resonance (ODMR) to magnetic fields. Furthermore, we demonstrate local temperature sensing of a microcircuit while simultaneously measuring an external magnetic field. Our results establish quantum emitters in hBN as a robust platform for multifunctional quantum sensing under realistic operating conditions.*

**Keywords**

*Thermometry, Magnetometry, ODMR, hBN, ZPL, Dual sensing*

Nanoscale quantum sensing is becoming an important aspect of many applications spanning chemical reactions, bio-imaging, and nanoelectronics(1–4). Amongst a variety of current sensing techniques(5–9), solid-state quantum emitters with optically addressable spins have attracted significant attention. These defects are often hosted by solid-state materials, which makes them compatible with a wide variety of applications(10–12).

Amongst the variety of color center hosts, defects in diamond, silicon carbide, and hexagonal boron nitride (hBN) have been extensively studied(13–21). A subset of defects in these hosts possesses a triplet ground state and exhibits a clear optically detected magnetic resonance (ODMR) signature. Consequently, the ODMR transitions can be utilized to sense external magnetic fields, electric fields, or temperature(12,22–24). However, simultaneous sensing using the same quantum sensor is impractical, requiring complex calibrations and measurement procedures (25–27), which limits their applicability in environments where both parameters fluctuate.
To address this challenge, one requires a narrowband quantum emitter that reacts independently to temperature and magnetic field. This can be achieved by monitoring the shift in the zero phonon line (ZPL) to gain information about the temperature, while independently monitoring the spin resonance to know the magnetic fields.
Figure 1 (a) illustrates the concept of dual sensing of magnetic field and temperature by combining ODMR and ZPL shifts from a single quantum emitter in hBN. The magnetometry system implemented in this work is controlled by the spin Hamiltonian of the optically active defect. For a typical spin-1 system, the spin Hamiltonian is shown in equation (1).

$$(1)\ \hat{H} = D(\hat{S}_z^2 - \frac{1}{3}S(S+1)) + E(\hat{S}_x^2 - \hat{S}_y^2) + g\mu_B B \cdot \hat{S}$$

The D and E terms are zero-field splitting (ZFS) parameters that come from the internal electron interaction and the structure of the lattice. The final term is the Zeeman interaction, which describes the coupling of an external magnetic field to the spin. As the ZFS parameters are dependent on the lattice structure, any thermal expansion of the lattice will change the resonance frequency. For spin-1 systems, the temperature and magnetic field cannot be separated without complex fitting and multiple calibrations.
On the other hand, for a spin-1/2 system, there is no electron pair to create the ZFS interaction, and the Hamiltonian is reduced to only the Zeeman interaction (equation (2)).

$$(2)\ \hat{H}_{Zee} = g\mu_B B \cdot \hat{S}$$

The two spin states $m_s = \pm 1/2$ split linearly with applied magnetic field, resulting in an ODMR resonance frequency that changes at the rate of the electron gyromagnetic ratio (~28 MHz /mT)(28). Without any contributions from the ZFS, the ODMR resonance frequency is temperature independent, making the spin 1/2 system ideal for magnetometry(29–33).
The thermometry sensing is based on the redshift of the ZPL position that is caused by the temperature dependence of the electronic bandgap. The ZPL wavelength shift is insensitive to Zeeman interactions created by the magnetic field(34). By exploiting these two independent parameters, simultaneous sensing of the magnetic field and temperature can be achieved using a single quantum emitter.
To test the multifunctional sensing capability, an hBN flake hosting a variety of ODMR active quantum emitters (see Methods for fabrication details) is placed on a heating block with a position-controlled magnet. The setup is shown schematically in Figure 1(b). Figure 1 (c) shows the confocal photoluminescence (PL) map at room temperature (23 ℃), under a magnetic field of 73 mT. The yellow dotted line illustrates the edge of the hBN flake. The bright spot marked by the red circle is the quantum emitter with a narrow emission peak, shown in Figure 1(d). The second-order correlation function ($g^{(2)}(\tau)$), shown in the inset, confirms that the emitter is a single-photon source. The emitter also exhibits an ODMR signal ~2.05 GHz, as shown in the supporting information (Figure S1).

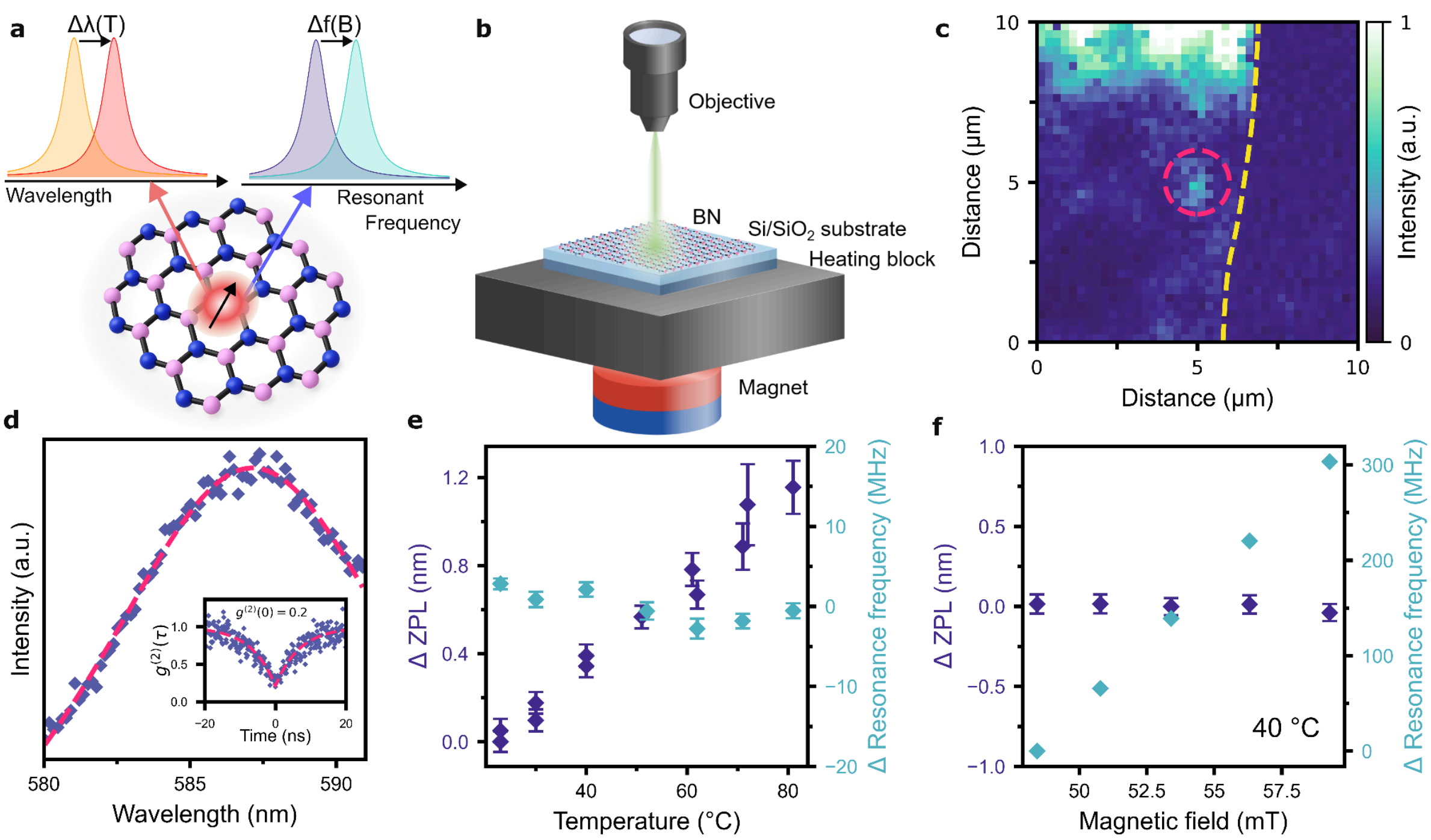

*Figure 1. Concept demonstration of simultaneous sensing of temperature and magnetic field using a quantum emitter in hBN. (a) A graphic depiction of a temperature-dependent PL spectra and a magnetic field-dependent ODMR resonant frequency. (b) Sample stage schematic with labeled components. (c) Normalized PL map showing the hBN flake edge marked with yellow, and an emitter marked by red. (d) The PL spectra of the emitter marked in (c), the emission has a wavelength of 587 nm and a ~15 nm FWHM. Inset: Autocorrelation of PL emission, with a g2(0) = 0.2. (e) Change in ZPL position (purple) and resonant frequency (cyan) as a function of temperature on the heating block. (f) Change in ZPL position (purple) and in resonant frequency (cyan) as a function of an external magnetic field, recorded at 40°C on the heating block.*

Next, the temperature-dependent ZPL peak position and ODMR resonance frequency of the emitter were characterized, as shown in Figure 1(e). With increasing temperature from 23°C to 80°C, the ZPL exhibited a redshift with a thermal response of 0.019 nm/°C (purple curve, figure 1(e)). This thermal shift is comparable to the previous study(35). In contrast, the ODMR resonance frequency remains nearly constant over the same temperature range, with a standard deviation of ~ 2.0 MHz, as shown in cyan in Figure 1(e).
The independence of the ZPL position from the magnetic field was verified by recording emission spectra under varying magnetic fields, as shown in Figure 1(f). Shifting the magnet in 0.5 mm steps over a 2 mm range, the ODMR resonance frequency was changed by ~300 MHz, corresponding to a magnetic field change of ~10.7 mT. This value agrees with the simulated magnetic field change of 10.9 mT (Figure S2). On the other hand, the ZPL peak position remained stable with a standard deviation of ~ 0.02 nm. These measurements were performed at a temperature of 40°C to demonstrate that both ZPL and ODMR signals can be read out simultaneously under the modified temperature and magnetic field. A corresponding room temperature dataset is provided in Figure S3. Combined, these results validate the independent responses of the ZPL and ODMR to temperature and magnetic field, enabling simultaneous dual sensing with a single emitter.

We now turn to study the dual sensing modality of the quantum emitters under realistic, on-chip conditions. For this purpose, we designed and fabricated a micro heating circuit to generate localized temperature variation on the emitters. The circuit was patterned near the hBN flake to facilitate controlled local temperature gradients on the emitters. A schematic of the procedure is shown in Figure 2(a).

Figure 2(b) shows the simulated temperature distribution (Joule heating) using COMSOL at an applied voltage of 1.7 V. The narrowest region at the center of the wire reaches a maximum temperature of approximately 73 °C, while a temperature gradient of 0.35°C/μm is observed over a distance of 50 μm from the wire edge. Figure 2(c) shows the voltage-dependent temperature profile extracted from the simulation. Two representative emitter positions were defined at distances of 7 μm (position A) and 14 μm (position B) from the wire edge along the y-direction, while centered along the x-direction. The simulation revealed a clear temperature difference between positions A and B, confirming that the circuit generates a spatially localized temperature gradient suitable for local temperature sensing.

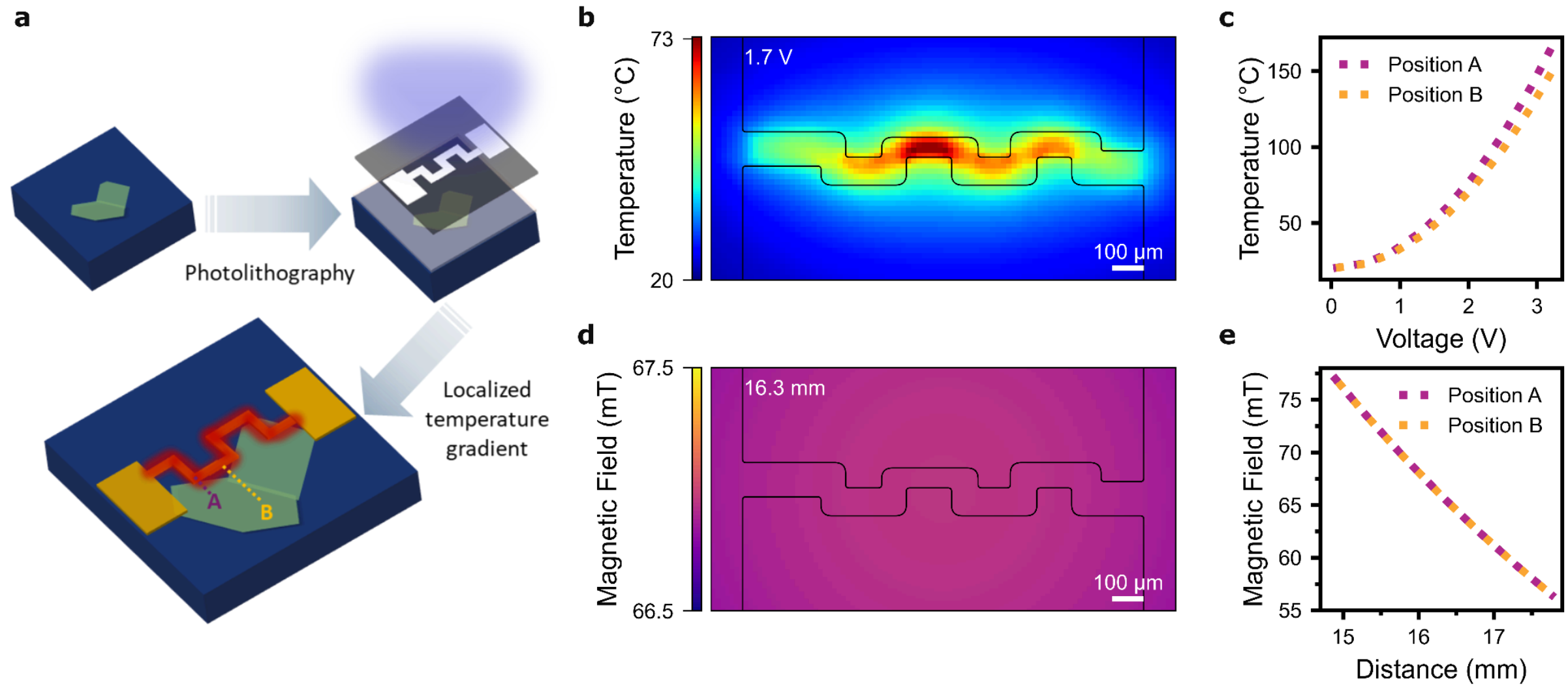

*Figure 2. Microscale heating circuit and resulting Joule-effect simulation. (a) Illustration of microscale heating circuit fabrication. Consisting of the flake selection followed by the photolithography step. Photoresist has been spun over the substrate and is exposed to patterned UV light. A voltage is applied to the circuit and a localized temperature gradient is created. The red line indicates the heating circuit. Position A (orange) and B (magenta) indicate emitter positions closer and farther from the heating region, respectively. (b) Simulated temperature distribution of the circuit heating at 1.7 volts. (c) Simulated temperature recorded from position A (orange) and position B (magenta) at distances of 7 and 14 μm, respectively. (d) Simulated magnetic field with a magnet placed 16.3 mm underneath the substrate. (e) Simulated magnetic field taken from the same positions in (c).*

Unlike the temperature distribution, the magnetic field is distributed uniformly across the chip area. Figure 2(d) shows the calculated magnetic field distribution when the magnet is located at the center and 16.3 mm from the substrate. The average strength of the magnetic field is ~66.9 mT. The magnetic field intensity as a function of the distance from the magnet is shown in Figure 2(e). The difference in magnetic field intensity between position A and B is below 10 µT within our observed area (<100µm×100µm), suggesting that the magnetic field is uniformly applied over the sample.

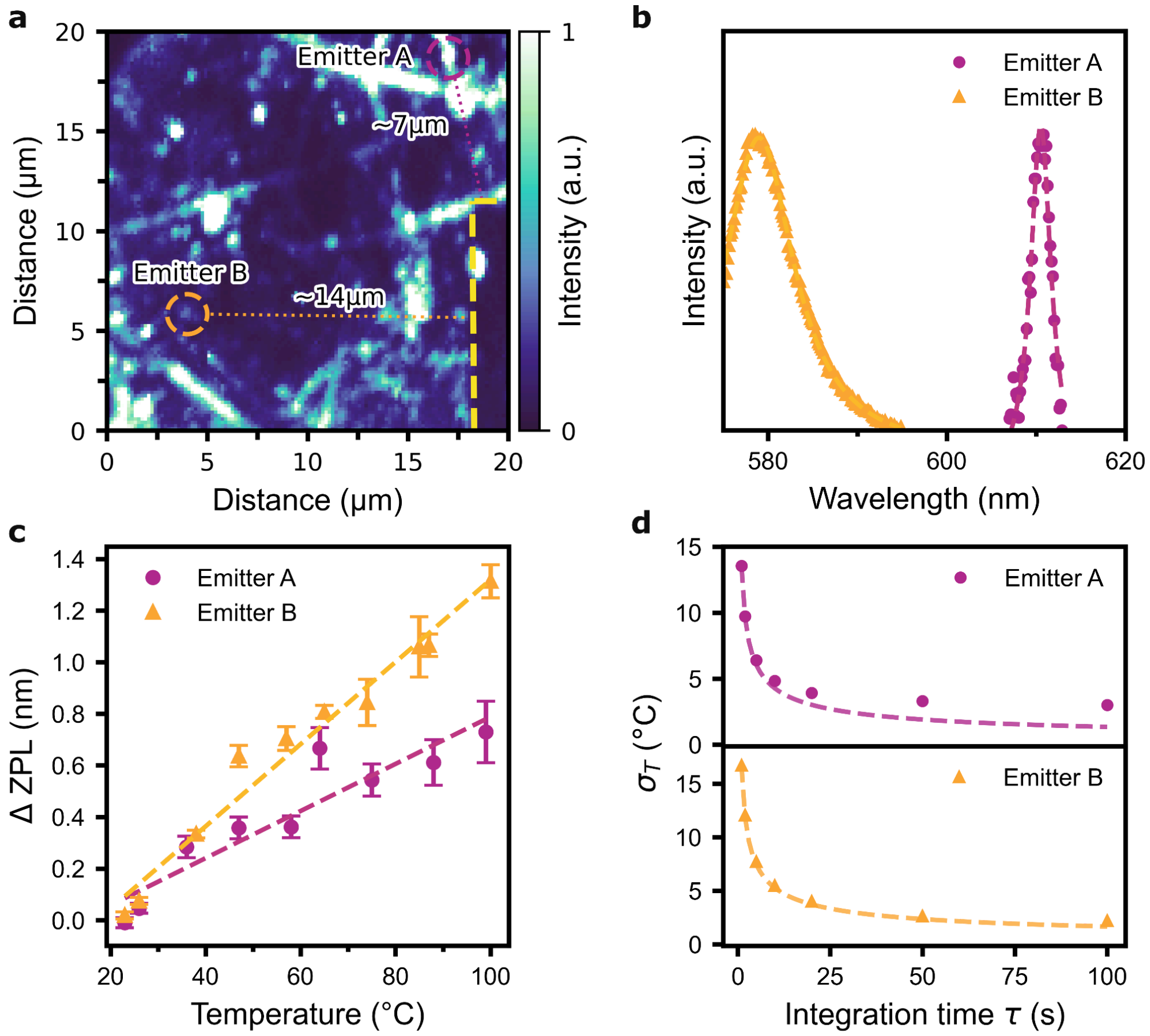

*Figure 3. Characterization of emitters used for simultaneous sensing. (a) Normalized PL map with hBN flake edge marked in yellow, Emitters A (magenta circle) and B (orange triangle) are marked with a circle. The area enclosed by the yellow lines is the location of the circuit. The distances from the wire to Emitter A and B are printed on the map and are ~7 and ~14 µm, respectively. (b) PL emission spectra for Emitters A (B) located at wavelengths 610 nm (579 nm), with a FWHM of ~2.8nm (~9.6 nm). (c) The change in ZPL position as a function of temperature on the heating block for Emitters A and B, with linear fits of 0.0092 nm/°C and 0.016 nm/°C, respectively. (d) Temperature uncertainty of the thermometer as a function of integration time. Dashed curve is a fitting curve of the shot noise limit* $1/\sqrt{N_{ph}}$.

To demonstrate the simultaneous temperature and magnetic field capabilities, two optically active emitters were identified near the heating circuit. Figure 3(a) shows a confocal map with Emitter A and B, located at distances of ~7 µm and ~14 µm from the edge of the heating circuit, respectively (marked by a yellow dotted line). Note that the positions A and B in the simulated result correspond to the positions of Emitter A and B from the circuit, respectively. The spectra of the emitters are shown in Figure 3 (b). The Emitters A (and B) exhibit a different full width at half maximum (FWHM) of 2.8 nm (and 9.6 nm), which are still narrow enough to carry the high-resolution temperature sensing. The calibration curve of the temperature dependent ZPL for both emitters is shown in Figure 3(c). The heating is performed using a thermal block heating element, as shown in Figure 1.

To assess the temperature sensing performance, we evaluated the uncertainty as a function of the integration time of the spectrum (Figure 3(d)). The uncertainties were extracted from the error of the Lorentzian fitting to the emission spectrum at each accumulation time. The extracted uncertainties of Emitter A and B are $\sim 13.4°C/\sqrt{Hz}$ and $\sim 16.4°C/\sqrt{Hz}$, respectively. The values are mainly limited by the signal-to-noise ratio of single-emitter measurements.

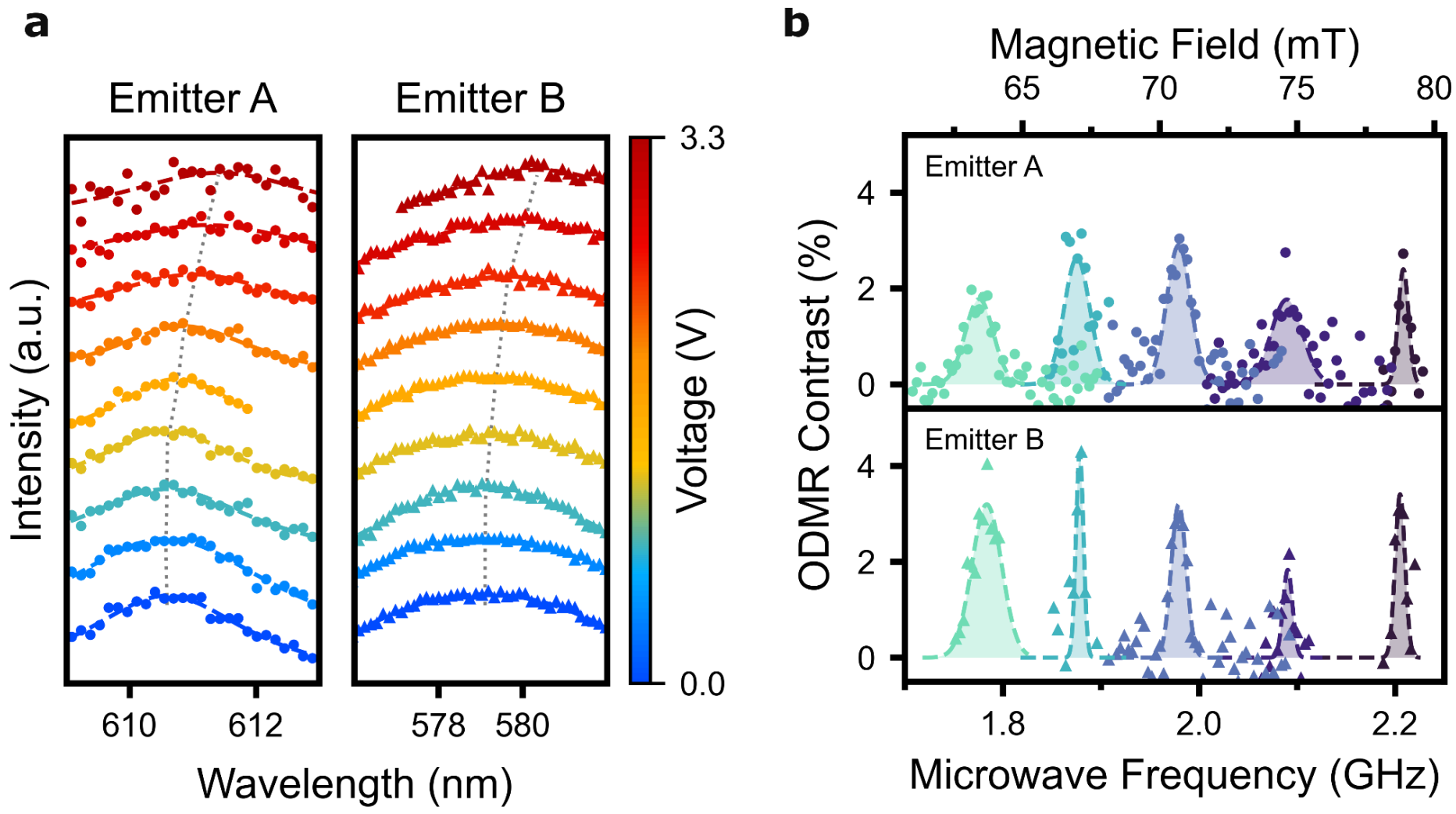

*Figure 4. Dual sensing characteristics of Emitters A and B. (a) Emission spectra for Emitters A and B over the applied voltage range of 0 to 3.3 V. The grey line marks the location of the fitted ZPL peak positions. (b) ODMR spectra for Emitters A and B recorded at different magnetic field strengths.*

The combined, simultaneous temperature and magnetic field sensing for Emitters A and B is shown in Figure 4. Both emitters' thermal response is depicted in Figure 4a. Over the applied voltage range of 0 V to 3.3 V, Emitters A (B) experienced ZPL redshifts of ~0.82 nm (~1.2 nm), corresponding to a local temperature of ~112°C (~101°C), respectively (Figure S4). The ODMR signal was independent from temperature with standard deviations of ~1.9MHz (~1.8MHz) across this temperature range (Figure S5(a)). The magnetic field response for both emitters is shown in Figure 4(b). As expected, both emitters experienced similar changes in the resonant frequency under the applied magnetic field. The magnetic field magnitude at each magnet position was calculated from the measured microwave resonance frequency. Emission spectra acquired at the same time confirmed the insensitivity of the ZPL emission against the applied magnetic field, with standard deviations of ~0.07 nm and ~0.06 nm for Emitters A and B, respectively (Figure S5(b)). The DC magnetic field sensitivity of Emitters

A and B was evaluated using a theoretical model (36), resulting in the calculated values of ~8.3 μT/√Hz and ~3.0 μT/√Hz, respectively.

We demonstrated on chip, dual sensing of temperature and magnetic field simultaneously using quantum emitters in hBN. We showed that the ZPL position is insensitive to magnetic field variations, while the ODMR resonance remains stable over the investigated temperature range. This result enables independent readout of the two physical quantities without cross-talk. The achieved sensitivities of $\sim 13.4°C/\sqrt{Hz}$ for thermometry, and $\sim 3.0\ \mu T/\sqrt{Hz}$ for magnetometry are promising for practical, real-world applications – especially in microelectronics. The intrinsic decoupling of temperature and magnetic field responses provides a significant advantage for practical sensing applications, where simultaneous and reliable multi-parameter measurements are required.
Further, while in our case a uniform magnetic field was applied, nanoscale variations of magnetic fields can be sensed using the same technique, down to the above-mentioned sensitivities. Additional parameters, such as fluorescence lifetime, can enable a complementary degree of freedom to sense chemical modifications, providing an all-around multifunctional quantum sensor. All in all, our results solidify the employment of quantum emitters in hBN as a promising platform for nanoscale sensing under realistic operating conditions.

## Methods

1. Sample preparation

Carbon-doped hBN was exfoliated onto diced 285 nm $SiO_2$/Si wafers using the scotch tape method. Visible emitters were activated by a high-temperature oxygen anneal. Using a Lindbergh blue tube furnace, the tube was connected to a scroll pump and purged with oxygen. The furnace was heated to 1000 °C with an oxygen flow rate of 1000 sccm and held for 4 hrs in this environment. The annealed substrates were placed in UV ozone for 4 hrs to remove any surface deposition and unstable emitters.

2. Fabrication of heating circuit

To design the micro-scale heating circuit, photoresist (AZ-1518) was deposited onto the prepared hBN substrate by spin-coating. Then, the sample was placed on PLANCK maskless UV lithography and exposed through a designed pixel array (1920x1080) to pattern a micro-scale heating circuit near the hBN flake. Then, we deposited Chromium using sputtering with a Dynavac CS 12-14 Dual Sputter System. Residual photoresist was removed by placing the sample in 40°C acetone for 1 hrs after deposition.

3. Optical characterization

PL measurements were performed using a home-built confocal setup that used a 532 nm CW laser and a x100 Olympus objective with a 0.9 numerical aperture. The surface of the sample was measured with a scanning mirror and a 568 nm long pass filter to block the laser falling on Excelitas avalanche photodiodes (APDs). An additional 617/14 nm band-pass filter was used for 'Emitter A' when performing autocorrelation and ODMR measurements. HBT measurements were recorded with two APDs, and coincidence was correlated by a PicoQuant time tagger.

4. Dual temperature and magnetic sensing
Global heating measurements were performed using a custom-built heating coil block, and the temperature was read out using a thermocouple and a multimeter. The power to the heating block and voltage were supplied to the heating micro circuit with a Tektronix 2623B system SourceMeter. CW ODMR was performed using an AnaPico APSIN4010 signal generator, the resulting frequencies were then amplified with a KeyLink KB0727M47C. A NdFeB magnet (N35) was placed below the sample on a z-axis controlled stage to provide a change in magnetic field.

5. Simulation
The Joule heating of the micro-circuit was simulated using COMSOL Multiphysics 6.3. A three-dimensional steady-state heat transfer model was employed. The applied current was determined from the experimentally applied voltage divided by the measured average resistance of the circuit. The temperature of the Si substrate and the surrounding air was set to 20 °C. Heat dissipation to the substrate and the surrounding air was included via thermal conduction. Material properties of chromium and silicon were taken from the COMSOL material library. The current was applied to a 1 mm × 1 mm chromium pad, matching the experimental configuration.
The magnetic field intensity distribution was calculated using a numerical analysis implemented in Python. The magnetic field was computed based on the Biot–Savart law by modeling the cylindrical magnet as a stack of current loops. The axial field of each loop was described by the standard expression, and the off-axis dependence was approximated using a distance-based formulation:

$$B_z(\rho, z) = (\mu_0 M R^2/2) \int_{z}^{Z+L} dz'/(R^2 + \rho^2 + z^2)^{3/2}$$

Where $\rho = \sqrt{x^2 + y^2}$ represents radial distance, $\mu_0$ is vacuum permeability, M is the magnetization, $R$ is the radius of the magnet, $L$ is the length of the magnet, and $z$ is the axial distance from the observation point.

**Acknowledgements**
The authors acknowledge financial support from the Australian Research Council (CE200100010, FT220100053, DP250100973) and the Air Force Office of Scientific Research (FA2386-25-1-4044).

**Conflicts of Interest**
The authors declare no conflicts of interest.

**Data Availability Statement**

The data that support the findings of this study are available from the corresponding authors upon reasonable request.

**References**

1. Degen CL, Reinhard F, Cappellaro P. Quantum sensing. Rev Mod Phys. 2017 Jul 25;89(3):035002. doi:10.1103/RevModPhys.89.035002

2. Atatüre M, Englund D, Vamivakas N, Lee SY, Wrachtrup J. Material platforms for spin-based photonic quantum technologies. Nat Rev Mater. 2018 May;3(5):38–51. doi:10.1038/s41578-018-0008-9

3. Wolfowicz G, Heremans FJ, Anderson CP, Kanai S, Seo H, Gali A, et al. Quantum guidelines for solid-state spin defects. Nat Rev Mater. 2021 Oct;6(10):906–25. doi:10.1038/s41578-021-00306-y

4. Awschalom DD, Hanson R, Wrachtrup J, Zhou BB. Quantum technologies with optically interfaced solid-state spins. Nat Photonics. 2018 Sep;12(9):516–27. doi:10.1038/s41566-018-0232-2

5. Fagaly RL. Superconducting quantum interference device instruments and applications. Rev Sci Instrum. 2006 Oct 11;77(10):101101. doi:10.1063/1.2354545

6. Sandell S, Chávez-Ángel E, El Sachat A, He J, Sotomayor Torres CM, Maire J. Thermoreflectance techniques and Raman thermometry for thermal property characterization of nanostructures. J Appl Phys. 2020 Oct 5;128(13):131101. doi:10.1063/5.0020239

7. Budker D, Romalis M. Optical magnetometry. Nat Phys. 2007 Apr;3(4):227–34. doi:10.1038/nphys566

8. Zhang Y, Zhu W, Hui F, Lanza M, Borca-Tasciuc T, Muñoz Rojo M. A Review on Principles and Applications of Scanning Thermal Microscopy (SThM). Adv Funct Mater. 2020;30(18):1900892. doi:10.1002/adfm.201900892

9. Balasubramanian G, Chan IY, Kolesov R, Al-Hmoud M, Tisler J, Shin C, et al. Nanoscale imaging magnetometry with diamond spins under ambient conditions. Nature. 2008 Oct;455(7213):648–51. doi:10.1038/nature07278

10. Du J, Shi F, Kong X, Jelezko F, Wrachtrup J. Single-molecule scale magnetic resonance spectroscopy using quantum diamond sensors. Rev Mod Phys. 2024 May 8;96(2):025001. doi:10.1103/RevModPhys.96.025001

11. Vaidya S, Gao X, Dikshit S, Aharonovich I, Li T. Quantum sensing and imaging with spin defects in hexagonal boron nitride. Adv Phys X. 2023 Dec 31;8(1):2206049. doi:10.1080/23746149.2023.2206049

12. Roberts H, Abudayyeh H, Li X, Li X. Quantum Sensing with Spin Defects Beyond Diamond. ACS Nano. 2025 Jul 1;19(25):22528–75. doi:10.1021/acsnano.5c00802

13. Rovny J, Gopalakrishnan S, Jayich ACB, Maletinsky P, Demler E, de Leon NP. Nanoscale diamond quantum sensors for many-body physics. Nat Rev Phys. 2024 Dec;6(12):753–68. doi:10.1038/s42254-024-00775-4

14. Castelletto S, Boretti A. Silicon carbide color centers for quantum applications. J Phys Photonics. 2020 Mar;2(2):022001. doi:10.1088/2515-7647/ab77a2

15. Ru S, An L, Liang H, Jiang Z, Li Z, Lyu X, et al. Room-Temperature Electrical Readout of Spin Defects in van der Waals Materials. Phys Rev Lett. 2025 Nov 24;135(22):220802. doi:10.1103/dlzw-dhsr

16. Qiao YF, Chen JQ, Zhou Y, Li PB, Gao WB. Pure spin squeezing of $h$-BN spins coupled to superconducting resonator. Phys Rev B. 2023 May 18;107(19):195425. doi:10.1103/PhysRevB.107.195425

17. Gong R, He G, Gao X, Ju P, Liu Z, Ye B, et al. Coherent dynamics of strongly interacting electronic spin defects in hexagonal boron nitride. Nat Commun. 2023 Jun 6;14(1):3299. doi:10.1038/s41467-023-39115-y

18. Gong R, Du X, Janzen E, Liu V, Liu Z, He G, et al. Isotope engineering for spin defects in van der Waals materials. Nat Commun. 2024 Jan 2;15(1):104. doi:10.1038/s41467-023-44494-3

19. Zhou J, Lu H, Chen D, Huang M, Yan GQ, Al-matouq F, et al. Sensing spin wave excitations by spin defects in few-layer-thick hexagonal boron nitride. Sci Adv. 2024 May;10(18):eadk8495. doi:10.1126/sciadv.adk8495

20. Huang M, Zhou J, Chen D, Lu H, McLaughlin NJ, Li S, et al. Wide field imaging of van der Waals ferromagnet Fe3GeTe2 by spin defects in hexagonal boron nitride. Nat Commun. 2022 Sep 13;13(1):5369. doi:10.1038/s41467-022-33016-2

21. Sun H, Yu P, Zhou X, Ye X, Wang M, Liu Z, et al. Room-temperature hybrid 2D-3D quantum spin system for enhanced magnetic sensing and many-body dynamics. Npj Quantum Inf. 2025 Dec 6;12(1):10. doi:10.1038/s41534-025-01152-4

22. Schirhagl R, Chang K, Loretz M, Degen CL. Nitrogen-Vacancy Centers in Diamond: Nanoscale Sensors for Physics and Biology. Annu Rev Phys Chem. 2014 Apr 1;65(Volume 65, 2014):83–105. doi:10.1146/annurev-physchem-040513-103659

23. Mu Z, Fraunié J, Durand A, Clément S, Finco A, Rouquette J, et al. Magnetic imaging under high pressure with a spin-based quantum sensor integrated in a van der Waals heterostructure. Nat Commun. 2025 Sep 29;16(1):8574. doi:10.1038/s41467-025-63580-2

24. Mu Z, Zhang Z, Fraunié J, Robert C, Seine G, Gil B, et al. Spin Defects in Hexagonal Boron Nitride as 2D Strain Sensors. Adv Funct Mater. 2026;36(36):e30638. doi:10.1002/adfm.202530638

25. Zhang Z, Gu K, Chen G, Sang L, Teraji T, Koide Y, et al. Highly Reliable Diamond MEMS Dual Sensor for Magnetic Fields and Temperatures with Self-Recognition Algorithms. Adv Mater Technol. 2024;9(13):2400153. doi:10.1002/admt.202400153

26. Shim JH, Lee SJ, Ghimire S, Hwang JI, Lee KG, Kim K, et al. Multiplexed Sensing of Magnetic Field and Temperature in Real Time Using a Nitrogen-Vacancy Ensemble in Diamond. Phys Rev Appl. 2022 Jan 7;17(1):014009. doi:10.1103/PhysRevApplied.17.014009

27. Hatano Y, Shin J, Nishitani D, Iwatsuka H, Masuyama Y, Sugiyama H, et al. Simultaneous thermometry and magnetometry using a fiber-coupled quantum diamond sensor. Appl Phys Lett. 2021 Jan 22;118(3):034001. doi:10.1063/5.0031502

28. Gottscholl A, Diez M, Soltamov V, Kasper C, Krauße D, Sperlich A, et al. Spin defects in hBN as promising temperature, pressure and magnetic field quantum sensors. Nat Commun. 2021 Jul 22;12(1):4480. doi:10.1038/s41467-021-24725-1

29. Scholten SC, Singh P, Healey AJ, Robertson IO, Haim G, Tan C, et al. Multi-species optically addressable spin defects in a van der Waals material. Nat Commun. 2024 Aug 7;15(1):6727. doi:10.1038/s41467-024-51129-8

30. Robertson IO, Whitefield B, Scholten SC, Singh P, Healey AJ, Reineck P, et al. A charge transfer mechanism for optically addressable solid-state spin pairs. Nat Phys. 2025 Dec;21(12):1981–7. doi:10.1038/s41567-025-03091-5

31. Chejanovsky N, Mukherjee A, Geng J, Chen YC, Kim Y, Denisenko A, et al. Single-spin resonance in a van der Waals embedded paramagnetic defect. Nat Mater. 2021 Aug;20(8):1079–84. doi:10.1038/s41563-021-00979-4

32. M. Gilardoni C, Eizagirre Barker S, Curtin CL, Fraser SA, Powell OFJ, Lewis DK, et al. A single spin in hexagonal boron nitride for vectorial quantum magnetometry. Nat Commun. 2025 May 28;16(1):4947. doi:10.1038/s41467-025-59642-0

33. Gao X, Vaidya S, Li K, Ge Z, Dikshit S, Zhang S, et al. Single nuclear spin detection and control in a van der Waals material. Nature. 2025 Jul 24;643(8073):943–9. doi:10.1038/s41586-025-09258-7

34. Li X, Shepard GD, Cupo A, Camporeale N, Shayan K, Luo Y, et al. Nonmagnetic Quantum Emitters in Boron Nitride with Ultranarrow and Sideband-Free Emission Spectra. ACS Nano. 2017 Jul 25;11(7):6652–60. doi:10.1021/acsnano.7b00638

35. Chen Y, Tran TN, Duong NMH, Li C, Toth M, Bradac C, et al. Optical Thermometry with Quantum Emitters in Hexagonal Boron Nitride. ACS Appl Mater Interfaces. 2020 Jun 3;12(22):25464–70. doi:10.1021/acsami.0c05735

36. Zhou F, Jiang Z, Liang H, Ru S, Bettiol AA, Gao W. DC Magnetic Field Sensitivity Optimization of Spin Defects in Hexagonal Boron Nitride. Nano Lett. 2023 Jul 12;23(13):6209–15. doi:10.1021/acs.nanolett.3c01881